\documentclass[a4paper,11pt]{article}
\usepackage{pos}

\usepackage{multirow}
\usepackage{longtable}
\usepackage{rotating}
\usepackage{pdflscape}
\usepackage{tabularx}
\usepackage{adjustbox}
\usepackage{floatrow}
\usepackage{lipsum}
\usepackage{hyperref}
\usepackage{amsmath}

\newcommand{\pine}{\texttt{PineForest}}
\newcommand{\coni}{\textsc{Coniferest}}

\title{Superluminous supernova search with \texttt{PineForest}}

\author[a]{Torsha~Majumder}
\author*[b]{Maria~V. Pruzhinskaya}
\author[c]{Emille~E. O. Ishida}
\author[d]{Konstantin~L. Malanchev}
\author[b,e]{Timofey~A. Semenikhin}
\author[f]{Vadim~V.~Krushinsky}

\affiliation[a]{University of Lethbridge,\\
4401 University Dr W, Lethbridge, Alberta AB T1K 3M4, Canada}

\affiliation[b]{Lomonosov Moscow State University, Sternberg astronomical institute,\\
Universitetsky pr. 13, 119234 Moscow, Russia}

\affiliation[c]{Université Clermont Auvergne, CNRS/IN2P3, LPCA,\\
F-63000 Clermont-Ferrand, France}

\affiliation[d]{McWilliams Center for Cosmology and Astrophysics, Carnegie Mellon University,\\
5000 Forbes Avenue, Pittsburgh, PA 15213, USA}

\affiliation[e]{Lomonosov Moscow State University, Faculty of Physics,\\
Leninskie Gory, 1-2, 119991 Moscow, Russia}

\affiliation[f]{Laboratory of Astrochemical Research, Ural Federal University,\\
ul. Mira d. 19, 620002 Yekaterinburg, Russia}

\emailAdd{torsha.majumder.work@gmail.com}
\emailAdd{pruzhinskaya@gmail.com}

\abstract{The advent of wide-field astronomical surveys has made available large and complex data sets. However, the process of discovery and interpretation of each potentially new astronomical source is, many times, still handcrafted. In this context, machine learning algorithms have emerged as a powerful tool to mine large data sets and lower the burden on the domain expert. Active learning strategies are specially good in this task. In this article, we used the \pine\  algorithm to search for superluminous supernova (SLSN) candidates in the Zwicky Transient Facility data releases. 
Starting from a data set of $\sim$14 million objects, and using eight previously confirmed SLSN light curves as priors, we found seven SLSN candidates. In the course of this search, we posted to the Transient Name Server three previously unreported transients: AT~2018moa and AT~2018mob as SLSN candidates, and AT~2018mpg as a possible supernova. We also discovered a re-brightening of the SLSN-I SN~2018fcg approximately 1300 days after its main burst, which poses an interesting problem regarding its origin. This study represents the first application of the \pine\ algorithm for the targeted search of rare astronomical transients, specifically SLSNe.}

\FullConference{Multifrequency Behaviour of High Energy Cosmic Sources - XV (MULTIF2025)\\
 9-14 June, 2025\\
Mondello, Palermo, Italy\\}

\begin{document}
\maketitle

\section{Introduction}
\label{introduction}

Large scale sky surveys, of which one of the most prolific  examples is the Zwicky Transient Facility \citep[ZTF, ][]{Bellm_2019}, have revolutionized the way astronomers interact with the data. Given that each data release (DR) contains a few billion light curves, visually inspecting the entire data set is not feasible. In this context, automatic machine learning algorithms offer the possibility to quickly classify large data sets \citep{2019arXiv190407248B}. However, such algorithms traditionally require a large number of previously labeled examples for training. This might not be a problem if the target of classification is a well established, and deeply scrutinized, class of astronomical objects. On the other hand, if the goal is to find rare or not well known objects, for which only a small spectroscopically confirmed sample is available, the task becomes much more difficult. 

Anomaly detection (AD) algorithms are design to automatically identify samples which deviate from the bulk of the data or those unlikely to be generated by the same underlying statistical distribution used to describe the overall behavior of a large data set \citep{xiong2010anomaly}. Such strategies have been extensively exploited in recent literature, and have proven to be successful when applied to a wide range of wavelengths and data formats in astronomy \citep[see e.g., ][]{2023ApJ...958..106F, 2023A&A...680A..74M, 2023ApJ...945..106P, 2023AJ....166..151P, 2024ApJ...974..172A, 2024SPIE13102E..12H}. However, given the peculiarities of astronomical data, such examples will frequently correspond to non-physical artifacts (noise, instrument failures, spurious detections due to weather conditions, among others).  Despite its clear success, AD applied to astronomical data is known to generate a significant fraction of non-physical outliers \citep{2021AJ....162..206S, 2021MNRAS.502.5147M}.

In order to identify scientifically interesting anomalies, feedback from the domain expert is paramount. This is where active learning \citep{settles2012active} strategies play an important role. Active learning is a class of machine learning algorithms where expert feedback is used on-the-fly to impose modifications in the training sample, or in the hyperparameters of  a particular machine learning model, thus allowing for incremental increase in performance. When used in the context of AD, these learning strategies allow the construction of personalized models, targeting the specific anomalies which satisfy the user's definition. Active AD algorithms like, Active Anomaly Discovery \citep[AAD, ][]{Das2017, 2021A&A...650A.195I} and Astronomaly \citep{2021A&C....3600481L}, have been applied to different astronomical data sets \citep[see e.g., ][]{, 2023A&A...672A.111P, 2024arXiv241001034A, 2024arXiv240910256S, 10.1093/mnras/stae2031} showcasing their potential in helping to select scientifically interesting objects which can posteriorly be scrutinized by the expert. 

Despite their undeniable success, all active learning strategies when applied to a new data set require at least a few iterations during which the expert needs to go through some obviously non-interesting candidates. This \emph{burn-in} period is necessary to start the fine tuning process of selecting candidates satisfying the user's requirements. Although this stage is unavoidable, it is possible to boost the convergence of the algorithm by providing examples of what the expert is searching for before the actual learning loop starts. In this work, we showcase one possible application of this strategy while searching for superluminous supernova (SLSN) in the ZTF DR8.  

Superluminous supernovae are transients significantly brighter than typical supernovae, with peak luminosities of $L_{\rm max} \gtrsim 10^{44}$ erg~s$^{-1}$. They are classified based on both their photometric and spectroscopic properties~\citep{2012Sci...337..927G,Gal_Yam_2019}. The first reports of their detection date from the turn of the century, however, they only became more frequent with the arrival of systematic time-domain sky surveys. Given their high luminosity, it is possible to detect them at relatively high redshifts, thus enabling their use in the study of star formation in the early Universe ($z\geq 2)$, the interstellar medium, or even use them for cosmological applications \citep{2018SSRv..214...59M}. 

Given that SLSNe are relatively rare and their consequent low incidence in ZTF DRs, they constitute a good science case to demonstrate how the use of already known examples can help boost results from active learning. We presents results from applying the \pine\ algorithm \citep{2025A&C....5200960K} 
enhanced with previously labeled examples. The characteristics of the data set are described in Section~\ref{sec:data} and our complete analysis pipeline is described in  Section~\ref{sec:method}. We present our findings in Section~\ref{sec:results} and conclusions in Section~\ref{sec:conclusions}.

\section{Data}
\label{sec:data}
The Zwicky Transient Facility \citep{Bellm_2019} constitutes a public, multi-band survey conducted in the $\{zg,zr,zi\}$ passbands, covering the entire northern sky with $\sim 2$ day cadence. It has a large instantaneous field of view CCD camera, which effectively leverages the entire focal plane ($\sim$47~deg$^2$) of the 48-inch Samuel Oschin Telescope at Palomar, resulting in a survey footprint which varies between 1000 and 2000 square degrees, depending on the season\footnote{\url{https://irsa.ipac.caltech.edu/data/ZTF/docs/releases/dr07/ztf_release_notes_dr07.pdf}}. ZTF  conducts bi-monthly public data releases, providing high-quality and reliable data products to enable time-domain science. In this work, we analyzed private and public data from  ZTF DR8, which spans from March 2018 to May 2020 ($58194\leq$ MJD $\leq58972$).

As a starting point, we used the data curated by \citet{10.1093/mnras/stae2031}. We examined only the clear $(catflags=0)$ $zr$-band light curves containing at least 300 detections each. Moreover, we considered only sources with galactic latitude $b\ge 30^{\circ}$. This criterion aims to exclude galactic plane, then prioritizing extragalactic transients. 
In total, we analyzed 13\,761\,212 ZTF objects, labeled by unique identifiers (OIDs).

\subsection{Confirmed superluminous supernova sample}

Our initial SLSN sample was constructed using spectroscopically confirmed SLSN from Transient Name Server\footnote{\url{https://www.wis-tns.org/}} (TNS), resulting in 50 ZTF light curves. However, only 8 light curves (corresponding to 6 individual SLSNe) fulfilled our selection cuts and were available in the data set curated by \citet{10.1093/mnras/stae2031}. The surviving SLSNe are listed in Table \ref{tab1}. We used these as positive class priors initially given to the \pine\ algorithm as one of the parameters to search for new SLSN candidates. 

\begin{table}[]
    \centering
    \begin{tabular}{|l|l|p{3.0cm}|}
        \hline
        \textbf{TNS Name} & \textbf{Type} & \textbf{ZTF OID(s)} \\
        \hline
        SN 2018don & SLSN-I & \texttt{821202400019642}, \texttt{791213100013510}, \texttt{792216100000865} \\
        \hline
        SN 2018hxe & SLSN-II & \texttt{822210300019023} \\
        \hline
        SN 2018kyt & SLSN-I & \texttt{790212100011207} \\
        \hline
        SN 2018lng & SLSN-II & \texttt{789212300014493}\\
        \hline
        SN 2019aje & SLSN-II & \texttt{756207300001552} \\
        \hline
        SN 2019fdr & SLSN-II & \texttt{634210400021472} \\
        \hline  
    \end{tabular}
    \caption{The prior list comprises 8 confirmed superluminous supernova light curves  (corresponding to 6 individual SLSNe). SN~2018don is classified as an SLSN-I, and we referenced the three different ZTF OIDs for SN~2018don from distinct ZTF Field IDs -- 821, 791, and 792. We assembled all the ZTF OIDs in the prior list based on their availability in the ZTF DR8 feature file curated by \citet{10.1093/mnras/stae2031}.}
    \label{tab1}
\end{table}

\section{Methodology}
\label{sec:method}

In this study, to identify new SLSN candidates, we started our investigation using the features from \citet{10.1093/mnras/stae2031}. Features from the spectroscopically confirmed SLSNe were presented to the \pine\ algorithm as priors. Subsequently, we performed the traditional active learning loop using  \pine. 

\subsection{Feature set}

We used pre-processed data with 53 extracted features from the ZTF DR8~\citep{10.1093/mnras/stae2031}. These features were computed using the \texttt{light-curve}\footnote{\url{https://github.com/light-curve/light-curve-python}} package developed by \citet{2021MNRAS.502.5147M}.
The feature set encompasses both magnitude and flux properties, such as \texttt{Amplitude, Kurtosis}, and \texttt{Standard Deviation}. A complete list of the distribution of these features and the priors is illustrated in Figures \ref{fig-features-1}-\ref{fig-features-5} in \ref{features-img}.
This ensemble of features elucidates different aspects of the light curve shape, with the tails of feature distributions highlighting objects with less common light curve properties or potential outliers in our data set.

\subsection{PineForest}

We employed the \pine\ algorithm, a component of the \coni\footnote{\url{https://coniferest.snad.space/en/latest/}} package \citep{2025A&C....5200960K} in our SLSN search. \texttt{PineForest} is an adaptive learning mechanism that distinguishes itself from the traditional Isolation Forest (IF, \cite{4781136}) by eliminating the less influential trees. At each iteration, the object with higher anomaly score is shown to the expert, who is asked to reply YES/NO. If the expert replies YES, the object with second highest anomaly score is shown. If the feedback is NO, the algorithm proceeds to eliminate trees which attributed a high anomaly score to that particular sample. A new set of trees are generated and the scores are recalculated. The procedure continues until the maximum budget stipulated by the user, with each iteration fine tuning the set of surviving trees which reflects the feedback from the expert. This approach results in a personalized model that is less likely to assign high anomaly scores to objects not of primary interest to the expert. A description of the \pine\ algorithm is available in \citet{2025A&C....5200960K}.

The scoring mechanism of this algorithm is defined as, 
\begin{equation}
  y_j = 
  \begin{cases}
    \! 
    \begin{alignedat}{2}
      & 1, \text{ if } x_i \text{ is normal},
      \\
      & 0, \text{ if } x_i \text{ is unlabeled}, 
      \\
      & -1, \text{ if } x_i \text{ is an anomaly}
      
    \end{alignedat}
  \end{cases}
\end{equation}
\begin{equation}
score(T_i) = \sum_{j=1}^N y_j\cdot d(T_i(x_j)), \\
\end{equation}
where $T_i$ is the $i^{th}$ tree in the forest, $y_j$ is the label assigned to each sample $x_j$, and $d(T_i(\cdot))$ is the depth of the $x_j$ sample in the tree.

In this work, we also use one additional feature from the \coni\ package: the possibility to give examples of the positive class to be used as priors, before the beginning of the iterative learning loop. In this option, the algorithm takes a list of examples (Table \ref{tab1}) which have already been scrutinized and classified. At first, it grows random decision trees, then uses the provided labels to filter out trees whose resulting anomaly score do not agree with the reported label. This procedure is repeated sequentially for all examples and only after this the iterative loop starts to pose questions to the expert. This procedure takes advantage of legacy data to pre-train the forest and thus lower the number of loops necessary to start showing interesting examples to the expert.

\subsection{Visual inspection}

At each iteration, the \pine\ algorithm presents a candidate to the expert. In return, it expects a boolean answer regarding whether the ZTF OID is considered a scientifically interesting anomaly. The expert will then search for information to support their answer, which should be based on the light curve behavior and auxiliary data, such as literature reviews, cross-matching with established databases and catalogs, and assessments against theoretical models.

To facilitate this human-in-the-loop feedback, we utilized the \texttt{SNAD ZTF Viewer}\footnote{\url{https://ztf.snad.space}} 
 \cite{Malanchev_2023}, a web interface developed to support the domain expert analysis with ZTF DRs. Each object in the \texttt{Viewer} is represented as multi-band light curves, providing access to individual exposure images and the Aladin Sky Atlas \citep{boch2014astronomical,refId0}. Additionally, it provides cross-matching with various other catalogs and databases of variable stars and transients, including \texttt{SIMBAD} and \texttt{VizieR} databases \citep{2000A&AS..143....9W}, \texttt{AAVSO VSX} \citep{2015yCat....102027W}, \texttt{Pan-STARRS DR2} \citep{Flewelling_2020}, \texttt{Gaia DR3} \citep{refId2}, and \texttt{ZTF alert brokers}: ALeRCE\footnote{\url{https://alerce.science/}}  \citep{2021AJ....161..242F}, Antares\footnote{\url{https://antares.noirlab.edu/}}  \citep{2021AJ....161..107M} and Fink\footnote{\url{https://fink-portal.org/}}   \citep{2021MNRAS.501.3272M}.

\section{Results}
\label{sec:results}

\begin{figure}
\includegraphics[width=1\textwidth]{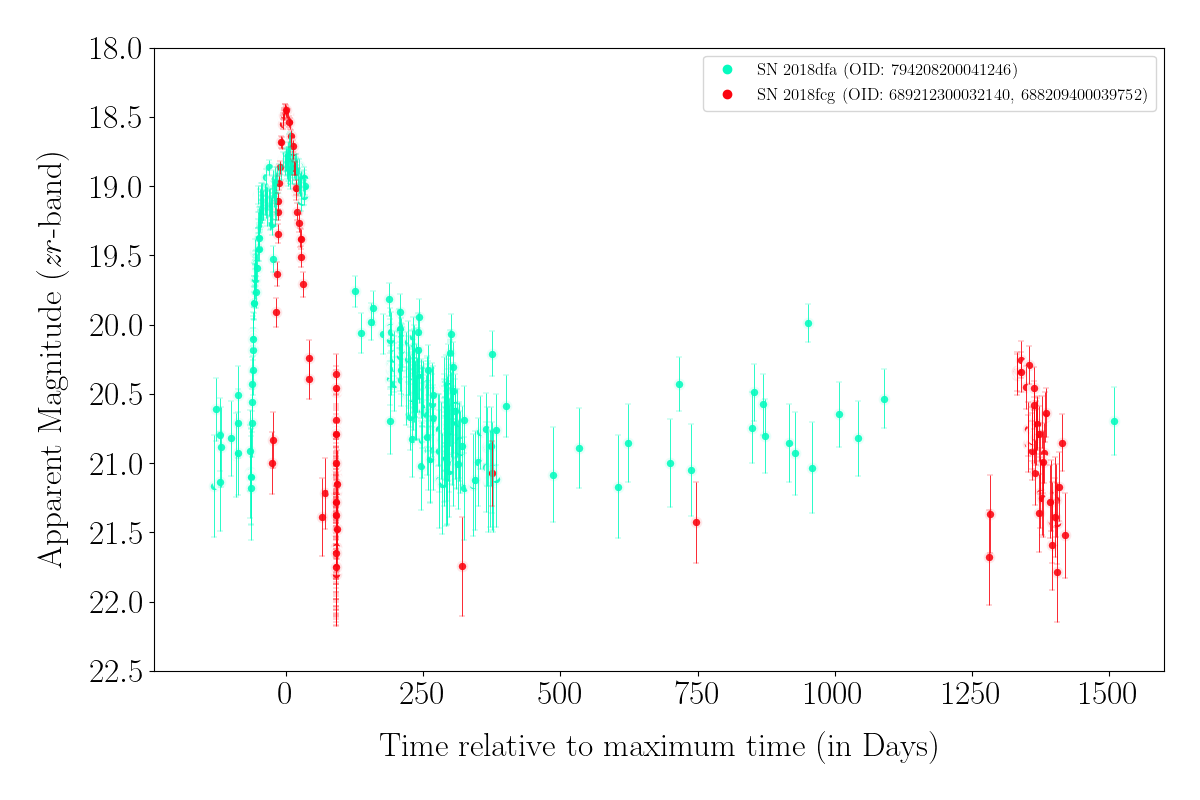}
    \caption{Superluminous supernova candidates with available spectroscopic redshift identified using the \pine\ algorithm. Only the ZTF OIDs included in the \pine\ outputs were used to construct the light curves, with the exception of SN 2018fcg, for which we used an additional \texttt{OID: 688209400039752} to show the rise of the late bump in its second peak. The MW extinction is not taken into account.
    }
    \label{spec}
\end{figure}

\begin{figure}
\includegraphics[width=1\textwidth]{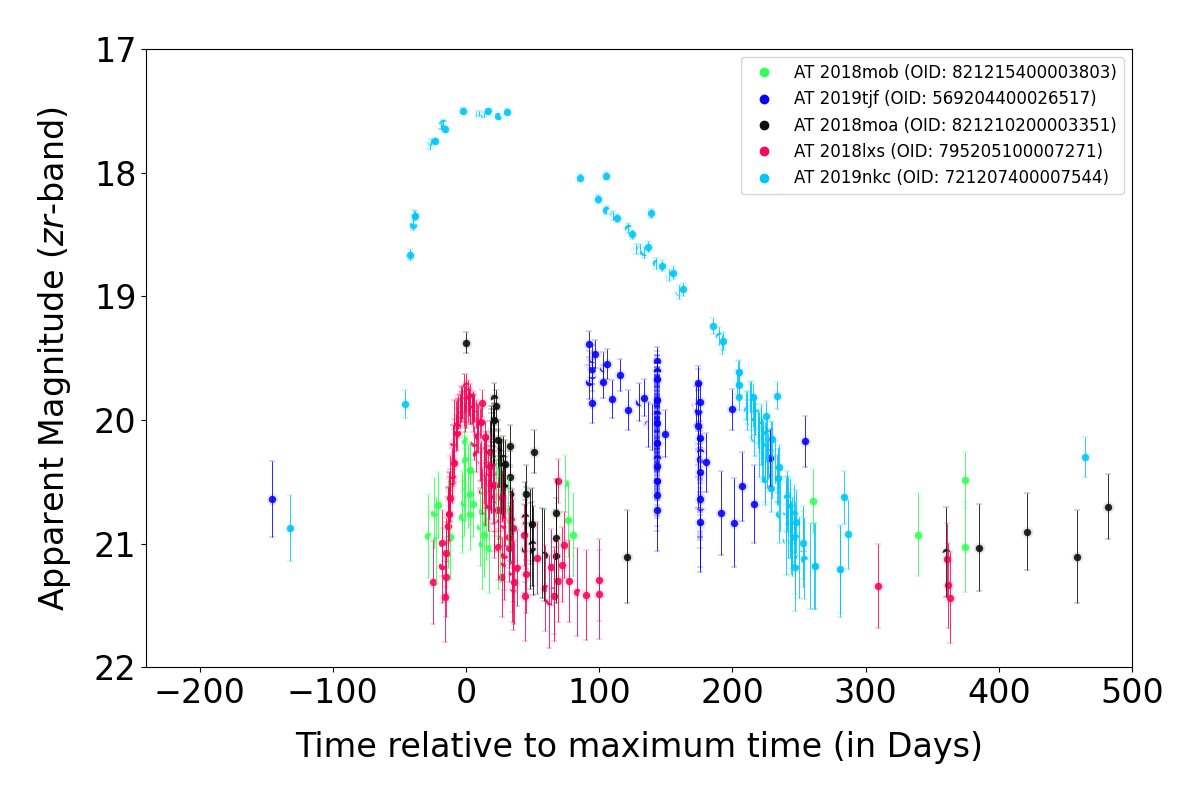}
    \caption{Superluminous supernova candidates with available photometric redshift identified using the \pine\ algorithm. Only the ZTF OIDs included in the \pine\ outputs were used to construct the light curves. The MW extinction is not taken into account.}
    \label{phot}
\end{figure}

We applied \pine\ to prioritize objects for human-in-the-loop inspection by domain experts. During the analysis, visual examination was performed, and objects showing features consistent with SLSNe were selected. 
Among the selected candidates, we identified 7 potential SLSNe, of which only one (SN~2018fcg) is a known, spectroscopically confirmed SLSN. Two objects have spectroscopic redshifts, while the remaining five have only photometric redshifts.
Additionally, SN~2018dfa was identified twice, associated with two different OIDs, \texttt{794208200041246} and \texttt{794208200022136}. Their $zr$-band light curves, after subtracting the host galaxy's contamination, are shown in Figures \ref{spec} and \ref{phot}. 

When selecting SLSN candidates, we adopted the commonly used threshold of peak absolute magnitude $M < -21$~mag in any band~\citep{2012Sci...337..927G}.
We employed Gaussian Process (GP) regression\footnote{\url{https://github.com/TorshaMajumder/PineSearch/tree/main/figures}} to model each light curve and estimate the peak apparent magnitude ($m_{r,~peak}^{GP}$) and the time of maximum ($t_{0,~peak}^{GP}$). The absolute magnitude was then calculated using the formula:

\begin{equation}
\label{eq:abs_mag}
M_{r,~peak}=m_{r,~peak}-5lgD+5-A_r,
\end{equation}
where $A_r$ is the Milky Way (MW) foreground extinction in the $r$-band ($A_r = 2.271\times E(B-V)$, \cite{Schlafly_2011}) and $D$ is the luminosity distance to the object in Mpc, assuming a flat $\Lambda$CDM cosmology ($H_0=70$ km~s$^{-1}$~Mpc$^{-1}$, $\Omega_m=0.30$). We did not apply any K-correction because the spectroscopic redshifts are relatively low, and the photometric redshifts have large uncertainties. 

Formally, 6 out of the 7 SLSN candidates have peak absolute magnitudes below the SLSN threshold of $-21$~mag. However, it is important to emphasize the large uncertainties associated with candidates having only photometric redshifts. Therefore, we also retained AT2018mob, whose estimated peak magnitude is $M_{r,~peak} = -20.94 \pm 0.07$~mag, consistent with the threshold within uncertainties. According to this definition, SN 2018dfa qualifies as a superluminous SN, although it is spectroscopically classified as an SN IIP in TNS~\citep{2018TNSCR1100....1F}.

The application of the \pine\ active learning pipeline has facilitated the identification of 3 new supernova candidates, designated as SNAD247/AT~2018moa\footnote{\url{https://www.wis-tns.org/object/2018moa}}, SNAD248/AT~2018mob\footnote{\url{https://www.wis-tns.org/object/2018mob}}, and SNAD259/AT~2018mpg\footnote{\url{https://www.wis-tns.org/object/2018mpg}}, which have been formally reported to the TNS. AT~2018mob is missing in the official ZTF alert stream.

Table~\ref{oid-table} lists the summary of the 7 objects identified during our search. The first and second columns represent the TNS name and classification of each object. The third and fourth columns provides the equatorial coordinates, right ascension (RA) and declination (Dec), in degrees. The fifth column indicates the line-of-sight reddening within our galaxy, E(B-V)~\citep{Schlafly_2011}. The sixth column contains the spectroscopic or photometric redshifts of the objects. The seventh column presents the Pan-STARRS DR2~\citep{Flewelling_2020} or ZTF reference magnitude used to correct for the host-galaxy contamination. The eight column represents the approximate absolute $zr$ magnitude derived using Equation~\ref{eq:abs_mag}. The estimated GPs peak apparent magnitude ($m_{r,~peak}^{GP}$) without MW extinction correction and the time of maximum in MJD ($t_{0,~peak}^{GP}$) are listed in columns nine and ten. Column eleven lists all available ZTF OIDs for the $zg$, $zr$, and $zi$ bands, with the highlighted OIDs corresponding to the \pine\ outputs.

\subsection{SN 2018fcg -- ZTF18abmasep}

Our study has led us to identify an additional anomalous multi-peaked light curve, SN 2018fcg, classified as a hydrogen-poor superluminous supernovae (SLSN-I), as reported at TNS~\citep{2018TNSCR1100....1F}, which exhibits a peak magnitude of $M_{r, peak} = -21.62$~mag. However, the full light curve is characterized by two distinct outbursts; the second is significantly fainter in the $zr$ band and has not been announced before this work (see Figure~\ref{spec}).

\begin{figure}
    \includegraphics[width=\linewidth]{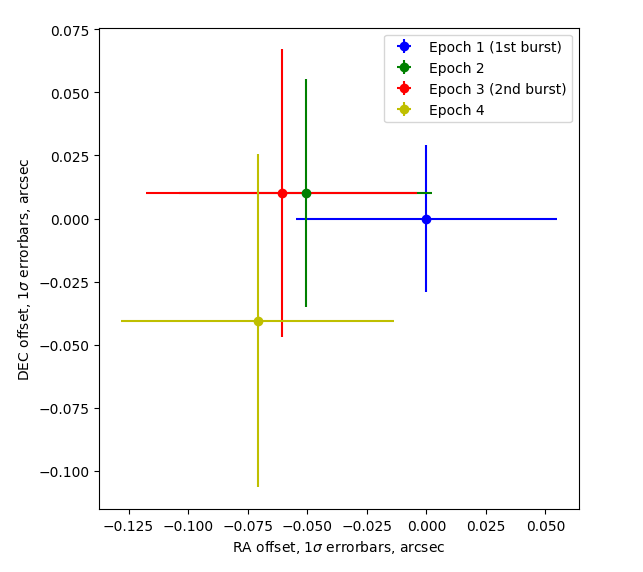}
    \caption{Measured positions of the second burst source and the host galaxy nucleus between and after bursts relative to first burst.
    }
    \label{offsets} 
\end{figure}

To investigate whether the two outbursts could be associated with different supernovae, we measured the positional offsets between the first and second bursts, as well as between the first burst and the host galaxy nucleus. For this purpose, we used stacked ZTF $zr$-band images obtained from 2018-08-13 to 2018-09-15 for the first burst (epoch 1), and from 2022-04-21 to 2022-06-08 for the second (epoch 3). The position of the host galaxy was derived from stacked images obtained during 2019-08-05 to 2021-09-05 (epoch 2) and 2022-07-24 to 2022-10-03 (epoch 4). To determine the position, 2D Gaussian fitting was applied. The measured offsets do not exceed the $1\sigma$ positional uncertainty (see Figure~\ref{offsets}), so no definitive conclusion can be drawn regarding the possibility of two separate supernovae.

Assuming the entire light curve originates from a single source, the prominent late-time bumps may be attributed to interaction with circumstellar medium (CSM), as indicated by \citet{Smith_2010}. However, the exceptionally long-lived nature of the light curve and the large time gap between the two peaks ($>1300$ days) make it difficult to interpret this behavior within standard SLSN scenarios and this event may represent a different kind of transient (see also~\cite{2025A&A...695A.142P}).

\subsection{Other outliers}
\begin{figure*}
    \includegraphics[width=\textwidth,height=0.40\textwidth]{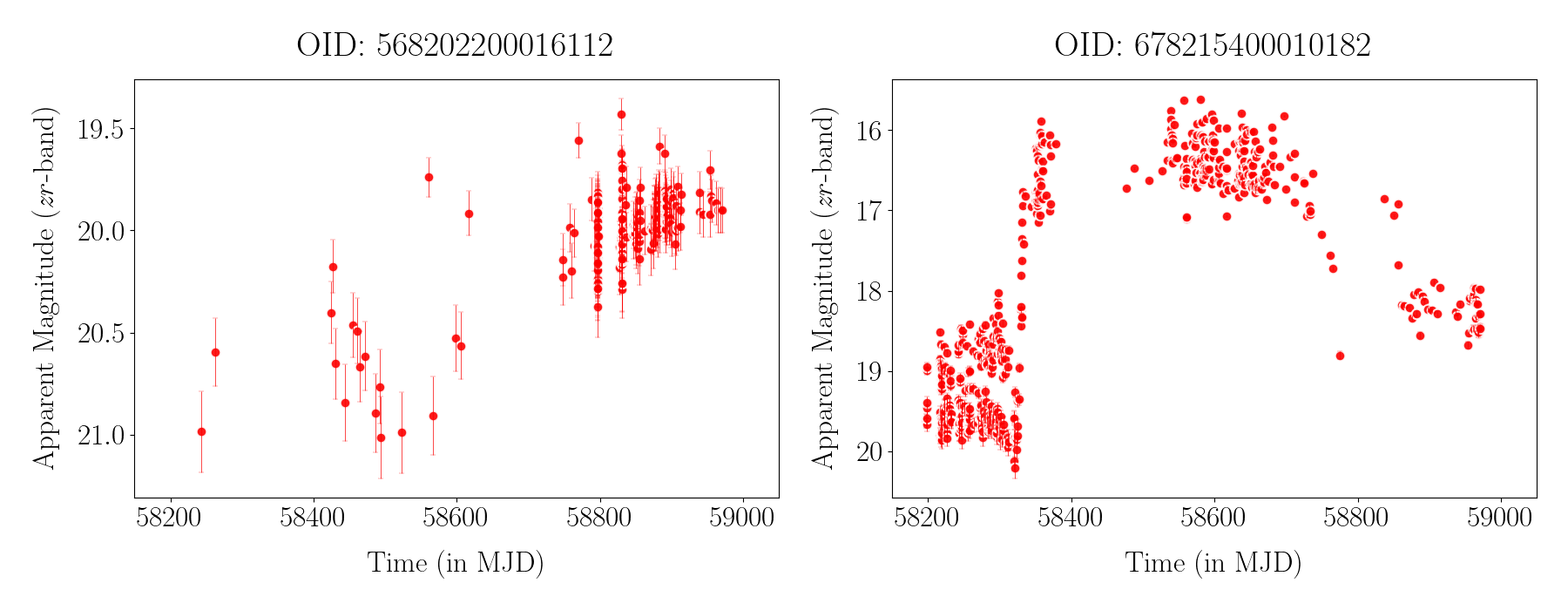}
    \caption{Outliers identified during the SLSN search using the \texttt{Pineforest} algorithm. The left image shows a QSO (\texttt{OID: 568202200016112})  and the right image shows AM Herculis-type variable BM CrB (\texttt{OID: 678215400010182}). The photometry corresponds to MJD range of $58194.0 \leq \text{MJD} \leq 59524.0$, indicating the specific data used to extract features for the \texttt{Pineforest} algorithm.
 }
    \label{byproducts} 
\end{figure*}

The majority of the \pine\  outputs were quasars (QSO) or QSO candidates, whose light curves can easily be mistaken for those of SLSNe. Additionally, the outliers included single stars, galaxies, artifacts, long-period variables, and known supernovae of classical types (Ia, II). Most of these objects exhibit broad light curves within the time range we examined. Figure~\ref{byproducts} illustrates examples, including a known QSO and an AM Herculis-type variable.

\section{Conclusions}
\label{sec:conclusions}

This study presents an application of the \pine\ algorithm within an active anomaly detection framework, specifically aimed at identifying unusual light curves in large-scale time-domain surveys. For the first time, \pine\ was applied to systematically search for superluminous supernova candidates within the Zwicky Transient Facility data.

To aid the convergence of the search, we incorporated prior knowledge in the form of a small set of known SLSN light curves as additional information. Using eight such light curves as priors, we identified seven SLSN candidates. We also posted to TNS three new transients: AT~2018moa, AT~2018mob, and AT~2018mpg. Furthermore, we discovered a re-brightening of SLSN-I, SN~2018fcg, which origin poses an interesting problem.

This strategy has the potential to yield further compelling discoveries regarding SLSNe once we use the entire prior catalog comprising 50 unique ZTF light curves for confirmed SLSN in the new data set curated for this research. Moreover, the methodology developed here can be adapted to a wide range of scientific applications beyond SLSN identification, highlighting the versatility and broader impact of integrating prior knowledge into machine learning workflows in astronomy and other scientific fields.

\section*{Acknowledgements}
We thank Petr Baklanov for the discussion on multi-peaked SLSNe. T. Semenikhin and M.~Pruzhinskaya acknowledge support from a Russian Science Foundation grant 24-22-00233, \url{https:
//rscf.ru/en/project/24-22-00233/}, for the analysis of SLSN candidates. Support was provided by Schmidt Sciences,
LLC. for K. Malanchev.

\appendix

\section{Table of SLSN candidates}\label{section:table}
We present here a complete list of superluminous supernova candidates found with the \pine\ active learning algorithm compiled with 8 SLSN light curves as priors present in the ZTF DR8.

\begin{landscape}
\small
\begin{longtable}{p{1.9cm}p{1.1cm}p{1.5cm}p{1.5cm}p{1.2cm}p{1.6cm}p{1.7cm}p{1.5cm}p{1.5cm}p{1.8cm}p{2.6cm}}
    \textbf{TNS Name} &
    \textbf{Type\footnotemark[13]}&
    \textbf{RA (deg)} &
    \textbf{Dec (deg)} &
    \textbf{E(B-V)} &
    $\mathcal{\mathbf{z}}$\footnotemark[14] &
    $\mathbf{m_{r, ref}}$ &
    $\mathbf{M_{r, peak}}$ &
    $\mathbf{m_{r, peak}^{GP}}$ & 
    $\mathbf{t_{0, peak}^{GP}~(MJD)}$ &
    \textbf{Notes\footnotemark[17]} \\
    \hline
    SN 2018dfa &SN IIP&230.21706&54.21550&0.01& 0.125\quad\quad\quad\small(sp) &\text{20.62}${\pm}$\text{0.02}\footnotemark[15]&\text{-21.72}${\pm}$\text{0.11} &\text{18.84}${\pm}$\text{0.11}&58350.4& $zr$~(\textbf{794208200041246}, 794208200022136, 793205100013372),  $zg$~(793105100016413, 794108200005886) $zi$~(793305100004986, 794308200006125, 793305100004986)\\
    SN 2018fcg &SLSN-I&317.40323&33.48323&0.17& \text{0.101}\quad\quad\quad\small(sp)&\text{22.16}${\pm}$\text{0.08}\footnotemark[15]&\text{-21.62}${\pm}$\text{0.01} &\text{18.44}${\pm}$\text{0.01}&58358.6& $zg$~(689112300050742, 1729101400031561), $zr$~(\textbf{689212300032140}, 688209400039752, 1729201400023934)\\
    AT 2019nkc &--& 235.19183 &38.84521&0.02&\text{0.109}$\pm$\text{0.064} (ph)&\text{20.81}${\pm}$\text{0.05}\footnotemark[16]&\text{-22.74}${\pm}$\text{0.17}&\text{17.50}${\pm}$\text{0.11}&58751.3&$zg$~(721107400003558), $zr$~(\textbf{721207400007544}), $zi$~(721307400020258)\\
    AT 2019tjf &--&135.74812&15.48116&0.04 &\text{0.207}$\pm$\text{0.092}\quad\quad\quad\small(ph)&\text{20.49}${\pm}$\text{0.06}\footnotemark[16]&\text{-21.02}${\pm}$\text{0.19}&\text{19.11}${\pm}$\text{0.16}&58655.0& $zg$~(518115100011724, 1564111400009400), $zr$~(\textbf{569204400026517}, 518215100006445, 1564211400013565), $zi$~(518315100017723, 569304400003869)\\
    AT 2018moa &--& 207.16852  & 63.57014 &0.02 &\text{0.429}$\pm$\text{0.381}\quad\quad\quad\small(ph) &\text{20.17}${\pm}$\text{0.02}\footnotemark[15]&\text{-24.13}${\pm}$\text{0.24}&\text{19.46}${\pm}$\text{0.14}& 58198.4 & $zg$~(821110200001558), $zr$~(\textbf{821210200003351}, 1852202100000983), $zi$~(821310200020112)\\
    AT 2018mob &--& 205.98618  & 64.60320 & 0.02 & \text{0.181}$\pm$\text{0.040} \small(ph) &\text{20.19}${\pm}$\text{0.02}\footnotemark[15]& \text{-20.94}${\pm}$\text{0.07} &\text{20.50}${\pm}$\text{0.05}&58245.0& $zg$~(821115400006169), $zr$~(\textbf{821215400003803}, 1852206300001446), $zi$~(821315400015085)\\
    AT 2018lxs &--& 252.30217  & 54.11179 & 0.02 & \text{0.293}$\pm$\text{0.122} \small(ph) &\text{21.39}${\pm}$\text{0.04}\footnotemark[15]& \text{-21.10}${\pm}$\text{0.09} & \text{19.84}${\pm}$\text{0.02} & 58287.1 & $zg$~(795105100002903, 796108200007512, 1798114400000405), $zr$~(\textbf{795205100007271}, 796208200014591, 1798214400007797), $zi$~(795305100008092, 796308200008245)\\\hline
\label{oid-table}
\end{longtable}
    \footnotetext[13]{Classification is taken from the TNS based on its availability.}
    \footnotetext[14]{All spectroscopic (sp) redshifts are obtained from the TNS, while photometric (ph) redshifts are sourced from~\url{https://www.legacysurvey.org/viewer/}.}
    \footnotetext[15]{Pan-STARRS DR2 magnitude~\citep{Flewelling_2020}.}
    \footnotetext[16]{ZTF reference magnitude.}
    \footnotetext[17]{The highlighted ZTF OIDs are obtained during the human-in-the-loop iterations.}
\end{landscape}

\section{Data Distribution of the Features Space and Priors}
\label{features-img}
\begin{figure}[htbp]
    \centering
    \includegraphics[width=\textwidth, height=0.90\textwidth]{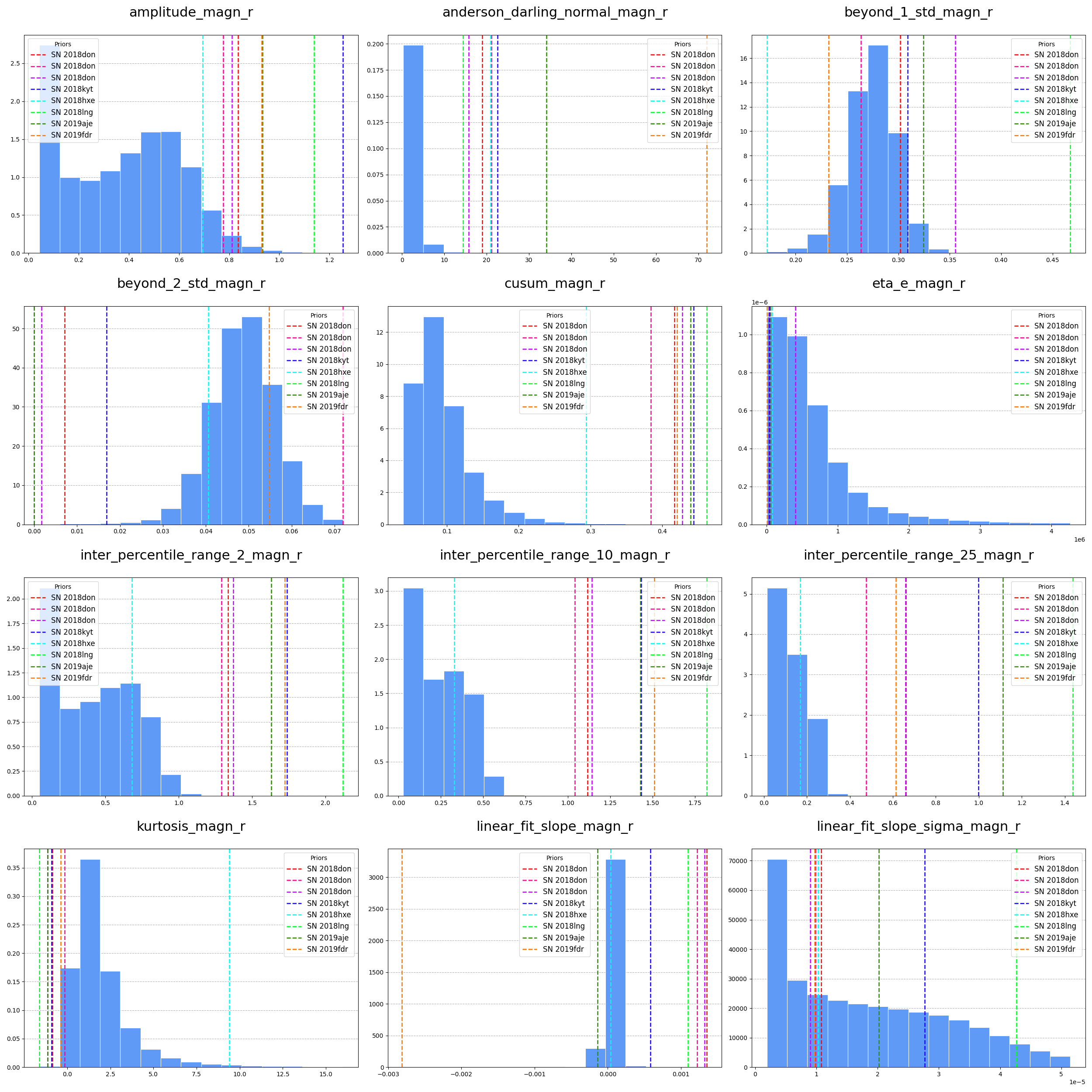}
    \caption{The feature space distribution from the ZTF DR8 alongside the eight priors detailed in Table \ref{tab1}. The features are extracted using the \texttt{light-curve} package and focus on various magnitude properties, including \texttt{Amplitude, Anderson-Darling Normal, Beyond N Std, Cusum, EtaE, Interpercentile Range, Kurtosis}, and \texttt{Linear Fit}. The 8 horizontal dashed lines represent the feature values for each light curves as priors.}
    \label{fig-features-1}
\end{figure}
\begin{figure}[htbp]
    \centering
    \includegraphics[width=\textwidth, height=0.90\textwidth]{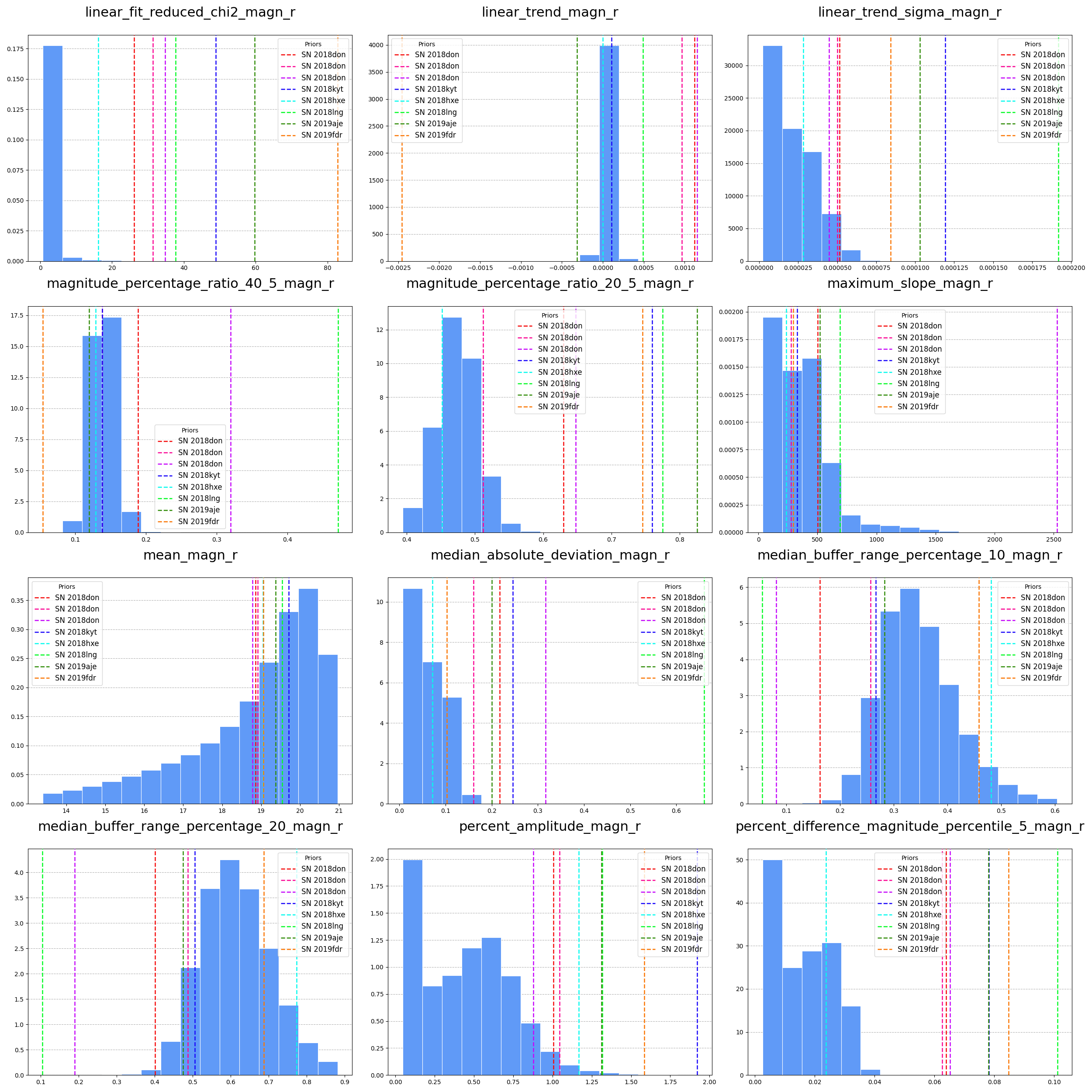}
    \caption{The feature space distribution from the ZTF DR8 alongside the eight priors detailed in Table \ref{tab1}. The features are extracted using the \texttt{light-curve} package and focus on various magnitude properties, including \texttt{Linear Fit, Linear Trend, MagnitudePercentageRatio, MaximumSlope, Mean, MedianAbsoluteDeviation, MedianBufferRangePercentage, PercentAmplitude}, and \texttt{PercentDifferenceMagnitudePercentile}. The 8 horizontal dashed lines represent the feature values for each light curves as priors.}
    \label{fig-features-2}
\end{figure}
\begin{figure}[htbp]
    \centering
    \includegraphics[width=\textwidth, height=0.90\textwidth]{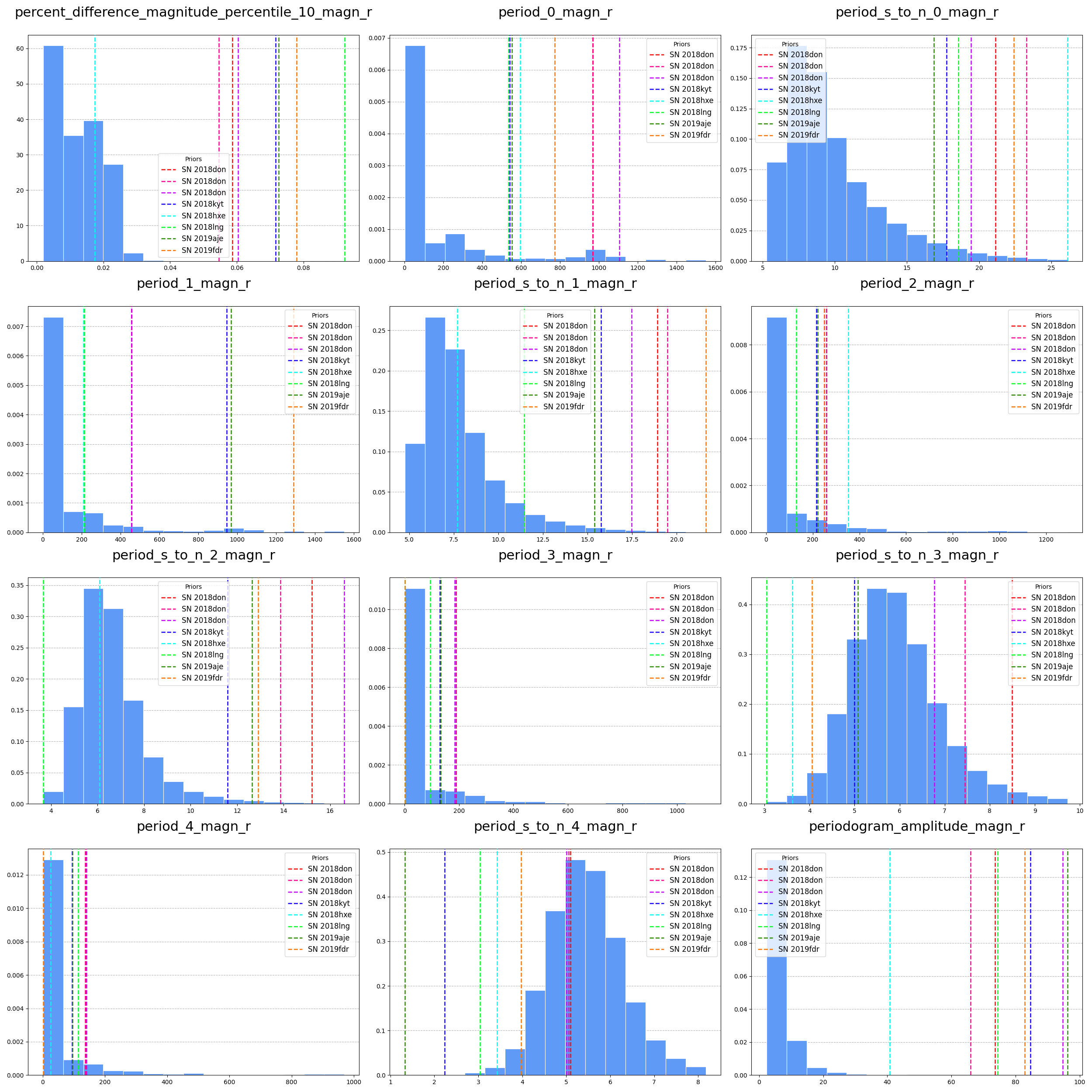}
    \caption{The feature space distribution from the ZTF DR8 alongside the eight priors detailed in Table \ref{tab1}. The features are extracted using the \texttt{light-curve} package and focus on various magnitude properties, including \texttt{PercentDifferenceMagnitudePercentile} and \texttt{Periodogram}. The 8 horizontal dashed lines represent the feature values for each light curves as priors.}
    \label{fig-features-3}
\end{figure}

\begin{figure}[htbp]
    \centering
    \includegraphics[width=\textwidth, height=0.90\textwidth]{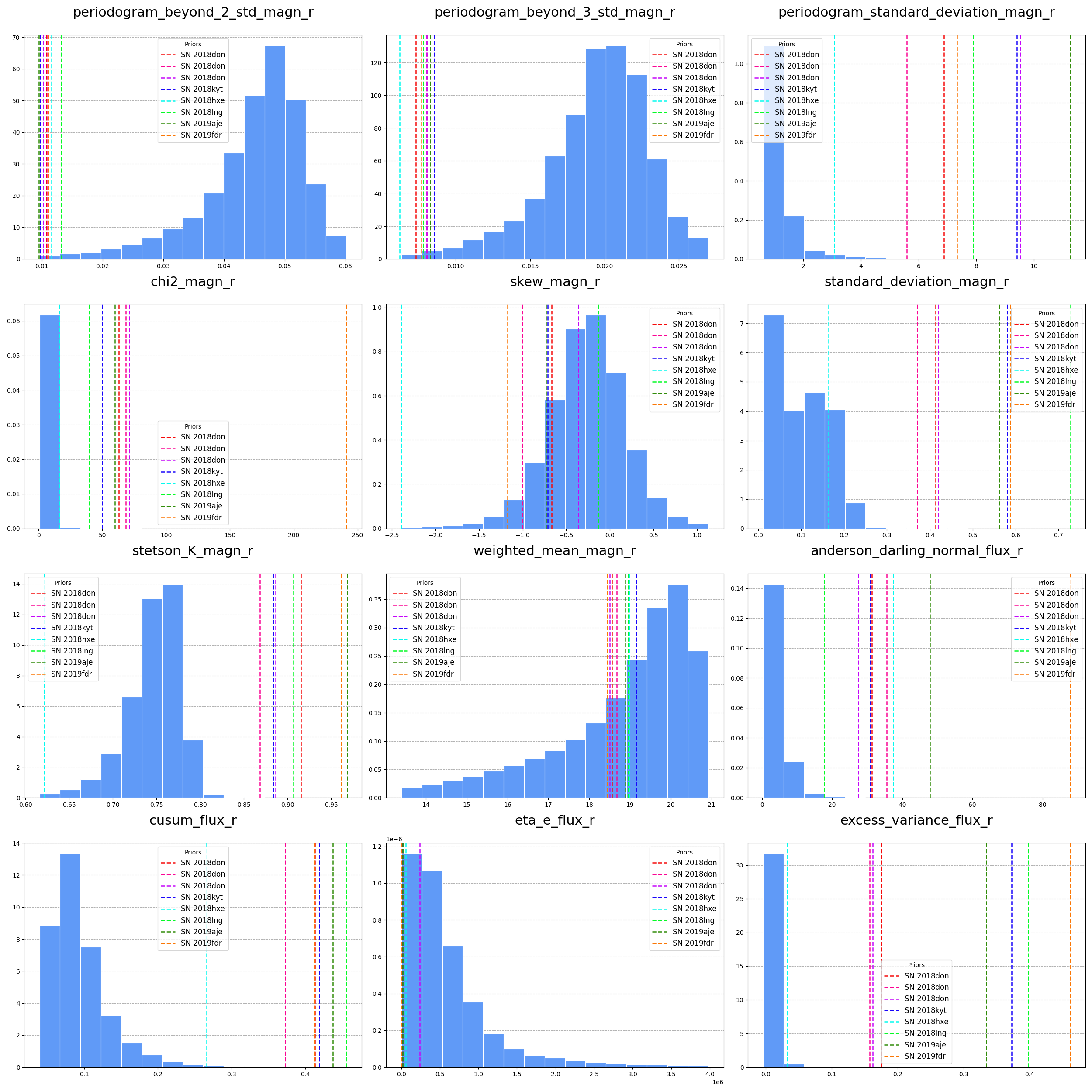}
    \caption{The feature space distribution from the ZTF DR8 alongside the eight priors detailed in Table \ref{tab1}. The features are extracted using the \texttt{light-curve} package and focus on various magnitude properties, including \texttt{Periodogram, ReducedChi2, Skew, StandardDeviation, StetsonK}, and \texttt{WeightedMean}. Additionally, it includes flux properties like \texttt{Anderson-Darling Normal, Cusum, EtaE}, and \texttt{ExcessVariance}. The 8 horizontal dashed lines represent the feature values for each light curves as priors.}
    \label{fig-features-4}
\end{figure}

\begin{figure}[htbp]
    \centering
    \includegraphics[width=\textwidth, height=0.90\textheight]{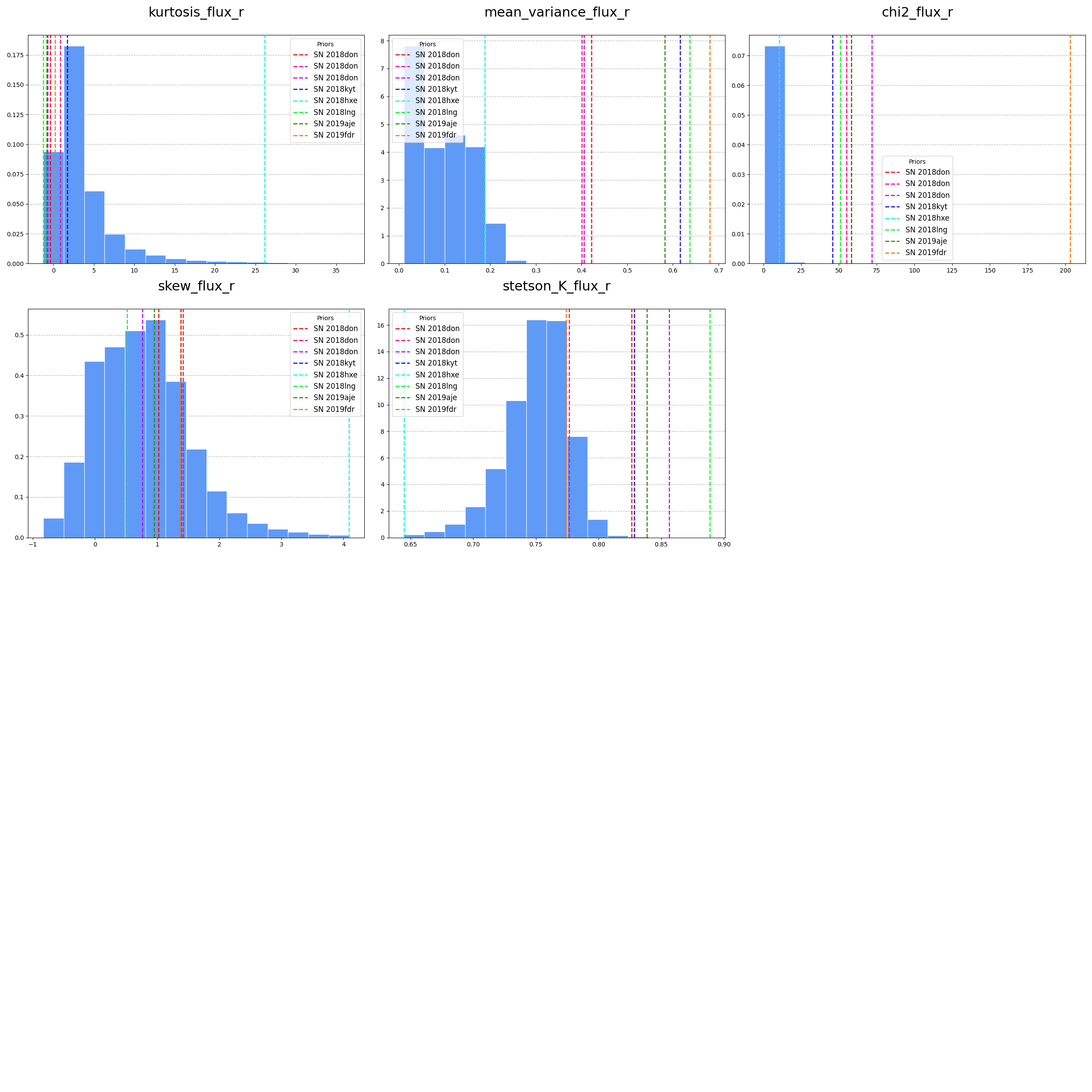}
    \caption{The feature space distribution from the ZTF DR8 alongside the eight priors detailed in Table \ref{tab1}. The features are extracted using the \texttt{light-curve} package and focus on various flux properties, including \texttt{Kurtosis, MeanVariance, ReducedChi2, Skew}, and \texttt{StetsonK}. The 8 horizontal dashed lines represent the feature values for each light curves as priors.}
    \label{fig-features-5}
\end{figure}

\clearpage
\bibliographystyle{elsarticle-harv} 
\bibliography{biblio}

\bigskip
\bigskip
\noindent {\bf DISCUSSION}

\bigskip
\noindent {\bf HARRISON COOK:} Have you looked at any of the $g-r$ color evolution for your ZTF sources?

\bigskip
\noindent {\bf MARIA PRUZHINSKAYA:} Yes, we have $g$, $r$, $i$ bands. The data are low quality so making color predictions is challenging but they all look normal for SNe.

\bigskip
\noindent {\bf MARIA DAINOTTI:} Do you have the plan to work on transients like gamma-ray bursts?

\bigskip
\noindent {\bf MARIA PRUZHINSKAYA:} Not for ZTF data releases. We checked, and only very few known afterglow observations are present in the data releases.

\bigskip
\noindent {\bf CARLO FERRIGNO:} With hybrid unsupervised classification plus human read, how do you manage to deal with upcoming large datasets?

\bigskip
\noindent {\bf MARIA PRUZHINSKAYA:} For anomaly detection human validation is always required, so the number of events an expert can realistically inspect remains roughly the same, even though the data volumes keep growing. This is exactly why we are developing active anomaly detection methods: with a limited number of expert labels or feedbacks, the model learns to focus on the most relevant candidates, making the whole process much more efficient.

\end{document}